\documentstyle[aps,preprint]{revtex}
\begin{document}
\title{Dynamical Phase Transition in a Driven Disordered Vortex Lattice}
\author{Seungoh Ryu$^1$, M. Hellerqvist$^2$, S. Doniach$^2$, 
A. Kapitulnik$^2$, D. Stroud$^1$}
\address{$^{(1)}$Dept.~of Physics, Ohio State University, Columbus, OH 43021
\\ $^{(2)}$Dept.~of Applied Physics, Stanford University, Stanford, CA 94305}
\date{\today}
\maketitle
\begin{abstract}
Using Langevin dynamics, we have investigated the dynamics of vortices 
in a disordered two dimensional superconductor subjected to a uniform driving 
current.   The results provide direct numerical evidence for a
dynamical phase transition between a plastic flow regime and a moving
``hexatic glass."   The simulated current-voltage characteristics are in
excellent agreement with recent transport measurements on
amorphous ${\rm Mo_{77}Ge_{23}}$ thin film superconductors.
\end{abstract} 
%74.60.Ec Mixed state, critical fields, and surface sheath
%74.60.Ge Flux pinning; flux creep, and flux-line lattice dynamics
%74.60.Jg Critical currents
%05.70.Fh Phase transitions: general aspects
%64.60.Cn Order-disorder transformations
While the effects of disorder on the
{\em static} Abrikosov lattice have been widely 
discussed\cite{larkin,statics}, 
the {\em driven} disordered lattice has been much less studied, 
despite its obvious relevance to typical transport measurements.   
Schmid and Hauger\cite{schmid73} analyzed interactions
between a pinning potential and the elastically deformed
moving lattice. Several groups\cite{jensen88,shi91} have
numerically demonstrated the importance of plastic deformation and 
associated dislocations in driven two-dimensional (2D) lattices. More 
recently, hysteretic, filamentary flow has been found numerically
in the plastically deformed regime\cite{jensen96}.

An actual {\em dynamic phase transition} between the
plastically deformed phase and a moving lattice for the vortex lines 
was first proposed and
numerically demonstrated in 2D by Koshelev
and Vinokur\cite{koshelev94}.  Their idea is that
the random ``pinning noise'' diminishes with increasing vortex velocity, 
allowing a high-velocity re-ordering.   Because of the disorder, the 
moving lattice has been argued to be a novel ``moving glass" phase with a 
finite transverse critical current\cite{giamarchi96}.
Similar dynamical ordering phenomena may be relevant
to Wigner electron crystals\cite{cha94}, vortex lattices in
Josephson arrays\cite{falo90}, magnetic bubble
arrays\cite{seshadri92}, and charge-density-wave 
systems\cite{fisher85,balents95}.  

Experimentally, a dynamical transition into a moving lattice phase has been
suggested by transport measurements\cite{shobo95}, by neutron 
diffraction\cite{yaron95} on flux lines in
2H-NbSe$_2$, and by the current-voltage (IV) characteristics in
the 2D amorphous superconductor ${\rm Mo_{77}Ge_{23}}$\cite{monica96}.  
The latter displays maxima in $dV/dI$\cite{monica96}, which may indicate
a dynamical ordering transition.

This Letter reports numerical studies of a model 2D vortex system which
show evidence for {\em three distinct dynamical phases}.  
The evidence comes from
calculations of both the $IV$'s and the translational and hexatic
order parameters.  We also obtain a finite transverse
critical current in the high-drive ordered phase, and a striking crossing
of all the $dV/dI$ curves for different temperatures at a single
point (I$_{cr}$, V$_{cr}$), in excellent agreement with experiment\cite{monica96}.

We consider $N_v\,$ two-dimensional 
vortices located at ${\bf r}_{i}$ in a thin superconducting
layer of thickness $d$, bulk penetration depth 
$\lambda(T) = \lambda(0) /\sqrt{ 1 - T/T_c}$, and fixed vortex density 
$n_B\equiv 1 / a_B^2 \equiv B /\phi_0,$ where $\phi_0 = hc/2e$. 
The classical Hamiltonian for the system is taken as
\begin{equation}
\label{eqaction}
{\cal H} = \big\{ \sum_{i\neq j} U \left(\frac{|{\bf r}_{i} - {\bf
r}_{j} |} {\lambda_{\perp}(T)}\right)
+ \sum_{i} V \left( {\bf r}_{i} \right) \big\}.
\end{equation}
The repulsive intervortex interaction is
$U(x) = {d \phi_0^2 \over 8\pi^2 \lambda(T)^2}K_0^* (x)$, where
$K_0^*(x)$ is a summation of the modified Bessel function 
$ K_0(x)$ over image vortices  
under periodic boundary conditions in the xy-plane\cite{ryu96}, 
and $\lambda_{\perp}(T)=\lambda^2(T) / d.$ 
The disorder potential $V(r)$ is taken as
$N_p\,$ potential wells of uniform depth 
$U_p(T,B) = \alpha_p \phi_0^2 /[  64 \pi^2 \lambda_{\perp}(T)]$ and radius 
$r_p$. 
To carry out the calculation for a given pin configuration, 
we first anneal the system from $T\sim T_c$ down to $T= 100 mK$, in
steps of $\triangle T = 200 mK$, using
the Metropolis algorithm\cite{ryu92}. At each temperature $T$ of interest,
we take a snapshot 
$\{ {\bf r_i} \}^0_{T,J=0}$ of the
vortex configuration during the annealing process, to be used as the initial
dynamical configuration.

The vortex dynamics are obtained from the overdamped equation
\begin{equation}
\label{eqlang}
\eta \dot{\bf r}_{i} (t) = {\bf f}_{i}^T(t) + {\bf f}_{i}^U(t) + 
 {\phi_0\over c} {\bf J} \times \hat{z} + {\bf f}^P ({\bf r}_{i} ).
\end{equation}
The first term on the right-hand side is the Brownian force
due to Gaussian noise\cite{koshelev94}.
The remaining terms are the forces due respectively to the other vortices, 
the applied current, and an {\em edge-smoothed} version of the
random pinning potential of eq.~\ref{eqaction}, as  described in detail
in\cite{ryu96}.  
In actual runs, we used the ${\rm
Mo_{77}Ge_{23}}$\cite{monica96} parameters
$\lambda(0) = 7700$ $\AA$, Ginzburg-Landau parameter 
$\kappa = 140$, $d = 60$ $\AA$ (and hence $\lambda_{\perp} = 98.8\mu {\rm m}$), 
$T_c = 5.63^\circ K.$   
Our bounding box containing 256 vortices for $B = 0.5 kG$ has an edge  $L_x \sim 3.2 \mu {\rm m}$, 
and we include up to $200\times 230$ image
boxes within a circle, comprising a total linear dimension 
$\tilde{L} \sim 6.5 \lambda_\perp(0)$. 
We use a time step of $0.1\sim1\times t_0$ where
$t_0  = {d \eta\lambda_\perp(0)\pi^2\over 16 B \phi_0}$,  an areal pin
density $B_p / B = 4.0$, and we take $r_p= 2\xi_{ab}(0)$, $\alpha_p =
1$. We also introduce a current density
scale $J_p$ by ${d\phi_0 J_{p}\over c} \equiv  U_p(0) / r_p$.  
For our parameters, $J_p = 5.05 \times 10^5 (A/cm^2)$, 
about 8 \% of the depairing critical current density.
Typically we discard the first $5000 t_0$ of time steps  
to assure achievement of a steady state.
 
To trace the disclination density $n_d(t)$ (i.~e., sites with coordination
number $\neq 6$), we perform Delaunay
triangulation on $\{ {\bf r_i}(t) \}_{T,J}$ every $10 t_0.$ 
We also measure (typically averaged over $\tau = 20000t_0$) the vortex
density-density correlation function
$G({\bf r}) = {1 \over \tau} \int^\tau dt <\rho({\bf r},t)\rho({\bf 0},t) >$, 
its Fourier transform $S({\bf q})$, and the hexatic order parameter
$\Psi_6$\cite{hexdef}. $IV$ characteristics are obtained from
the center of mass drift velocity of the entire vortex ensemble, and   
the Josephson relation ${\bf E} = n_B \phi_0 {\bf v}/ c\times \hat{z}.$

With no pins, we find that the lattice has a first order melting
transition at $T_m = 0.41 T_c$, with a Lindemann number of $0.19$, and a
simultaneous vanishing of both translational and hexatic order.  This
$T_m$ is reasonably close to the upper bound obtained from the
dislocation unbinding picture: $k_BT_m = A_1 (d\phi_0^2 / 128 \sqrt{3} \pi^3
\lambda^2(T_m) ) $, provided the renormalization constant 
$A_1 \approx 0.85$\cite{fisher80}.   With point pins,
the vortices freeze into a disordered state with some residual hexatic order.
An upper limit $T_f^+$
for the freezing temperature may be operationally defined as the point 
where the self-diffusion measure
$<|{\bf r}_i(t) - {\bf r}_i(0)|^2>$ converges to a t-independent value.
We find that the $T_f^+ \sim 2.0 K \sim 0.87 T_m.$

Fig.\ \ref{figone} shows the simulated $IV$ characteristics at several
temperatures.  The results at the lowest temperature $(T/T_m = 0.043)$ 
most clearly show that there are {\em three distinct regimes}\cite{pinnedlattice}.
(i) At low driving currents$(J/J_p < 0.2)$, 
the lattice is pinned and the voltage is generated mainly by
{\em rearrangements of topological defects}, which produce a very slow and
unsteady advance of the pinned lattice.
At least over the simulation time
($\sim 50,000 t_0$), the quenched-in hexatic order
stays robust; we may therefore call this regime a 
{\em pinned hexatic vortex glass} (PHVG). (ii) 
At moderate $J$ $(0.2<J/J_p<1.2)$, the $IV$'s are
strongly non-linear.  The hexatic order is progressively destroyed
with increasing $J$.  The nonlinearity originates from the nature of
plastic deformations in this regime, with highly filamentary 
vortex trajectories (cf.\ insets of Fig.\
\ref{figtwo}).   Deep into this state, $\Psi_6 \approx 0$
and the system is a
{\em plastic flow vortex liquid} (PFVL).  (iii) At high $J$, the lattice 
heals incompletely, and hexatic order is 
dramatically enhanced$(\Psi_6 \sim 0.6)$  over the static phase (i) 
$(\Psi_6 \sim 0.03)$.  Translational order is still disrupted by a
co-moving pattern of neutral disclination pairs.
Both g(r) and S(k) display a strong anisotropy. 
We may thus call this state a {\em moving
hexatic vortex glass} (MHVG).

The simulated $dV/dI$ curves shown in Figs.\ \ref{figtwo} and
\ref{figthree} reproduce most of the essential experimental 
features\cite{monica96}.
At the lowest $T$, Fig.\
\ref{figtwo} shows that  the peak in the calculated $dV/dI$ coincides
with the dip in $\Psi_6$.
At higher temperatures the peak broadens and merges with the
region at which $\Psi_6$ starts to increase again. 
In the simulations, the dynamic 
freezing transition (operationally defined as the sharp
increase of $\Psi_6$ at $J \sim 0.8 J_p$) falls on the 
right-hand shoulder of the peak.
In the experiment\cite{monica96}, the peak was interpreted to 
represent the increase in lattice correlation as it follows remarkably 
well the predicted form of Koshelev and Vinokur\cite{koshelev94}.
We therefore conclude that the peak in $dV/dI$
marks the point where lattice correlations {\em start to 
increase} in both orientational and positional order. 
In the plastic regime [left shoulder of the peak], 
the density and tortuosity of the calculated vortex 
trajectories are very sensitive to both $J$ and $T$ 
(presumably via the $T$-dependent effective shear modulus of the 
vortex lattice).  Qualitatively,
the current paths resemble percolation paths
just above the percolation threshold.  We suggest that
the current travels along an ``infinite connected cluster'' of open paths.
These paths occupy an area which increases with increasing $J$, perhaps
as a power law in (J-J$_0$), $J_0$ being the current density
where the infinite cluster first forms. 

As $T \rightarrow T_m$, the $dV/dI$ 
isotherms all tend to cross at a $J_{cr} \sim 0.12 J_p,$ strikingly
reproducing a prominent experimental feature\cite{monica96}. 
The pinned fraction of the vortices also starts a sharp 
decrease at this same value of $J/J_p$. 
Furthermore, as the inset of Fig.\
\ref{figthree} shows, this crossing coincides with the minimum in $\Psi_6$. 
For $J < J_{cr}$, we found that
the vortex flow is composed of two components: 
unpinned individual vortices in the flux-flow state; and 
pinned vortex islands with local hexatic order.  The latter
contribute to the flow through intermittent motion.
As $J$ increases toward $J_{cr}$, the size of the pinned islands shrinks and
consequently, the moving vortex trajectories become less tortuous.
We speculate that for each
$T$, $J_{cr}(T)$ is the current density where the $dV/dI$ for
the pinned and unpinned contributions become comparable.   
The reasons for the crossing of the $dV/dI$'s, and for the coincidence of
this crossing with a minimum in $\Psi_6$, will be discussed in greater detail 
elsewhere\cite{elsewhere}. 

Finally, we consider the MHVG.  In all our simulations, 
the moving healed lattice carries along 
bound disclination pairs ranging in density from $0.05$ at $T/T_m = 0.043$ 
to $0.4$ at $T/T_m =0.96$.   In the rest frame of
the average velocity $v$, the lattice has 
local triangular order, but long-range translational order
is disrupted by the disclination pairs (cf\/ Fig.\ 4).

This behavior suggests that the MHVG/PFVL is really dynamical
freezing, as suggested by \cite{koshelev94}.  To confirm this
interpretation, we can estimate the effective ``disorder temperature'' $T_d$.
Viewed in the moving frame, the pins
provide a random time-dependent potential which corresponds to an effective
$T_d$.  For the present pinning potential, a simple 
analysis, including the vector nature of the forces, yields
$k_BT_d = \frac{0.56 U_p^2}{va_pd\eta}.$ 
Here a$_p = 1/\sqrt{n_p}$,
where $n_p$ is number density of pins.  For our system, at zero temperature, 
$U_p = 7.5 \times 10^{-15}$ erg, and
$a_p = 1.02\times 10^{-5}$ cm.  
Using $E = Bv/c$, where $E$
is the electric field, we finally obtain
$T_d (K) \approx 6.6 \times 10^{-12}/[Et_0 {\rm (volt-sec/cm)}]$.
For low temperatures ($T/T_m \approx 0.043$), 
the PHVG/MHVG melting transition occurs near 
$E t_0 \approx 3 \times 10^{-12}$ volt-sec/cm, or equivalently 
$T_d \approx 2.2 K$, in excellent agreement with the simulated pin-free
melting temperature.

Note that in the MHVG phase,
the lattice principal axis is {\em not always parallel} to the 
driven direction. 
The degree of misalignment in the annealed lattice 
correlates with the angle between 
the drift and driving directions\cite{schmid73},
suggesting the possibility of an anomalous Hall effect.
This misalignment may occur because the lattice in 
MHVG is biased toward a direction which is frozen in during the 
$J=0$ annealing process, and thereafter 
frustrated from rotating parallel to the applied drive. 
To confirm this, we repeated with random initial configurations 
instead of an annealed state.
The lattice then oriented its principal axis
parallel to the driven direction, but
the resulting state was {\em higher in total energy than in
the annealed case, with about twice the number of disclinations}.

Finally, we note
the non-zero {\em transverse critical current} in MHVG.  
To see this effect, the lattice is first set in motion along a 
prescribed direction $\hat{\bf \ell}$. 
(of magnitude $J_{\ell}/J_p = 2.6$ in our simulations) for $20,000 t_0$. 
Then a small transverse current $0< J_{t}/J_{\ell} < 0.01$ is
turned on along $\hat{\bf t} = \hat{\bf \ell} \times \hat{z},$ and the
lattice transverse velocity $v_t$ 
is measured over $10,000 t_0.$  The results
are shown in Fig.\ \ref{figfour}
for two angles $\theta$ between $\hat{\bf \ell}$ and a local principal 
axes of the moving lattice: $\theta=0$ and $\theta = \pi/6$.
For $\theta = 0$ and $T/T_m = 0.043$, $v_t$ drops 
abruptly at $J_t/J_{\ell} = 0.002$.  For $\theta = \pi/6$, this drop is
much less sharp.  This transverse barrier seems to disappear very rapidly 
with increasing $T$, as shown  for $T/T_m = 0.43,$  $\theta = 0.$
A possible explanation for this behavior is suggested by the inset,
which shows the quenched disclination patterns 
(open circle and squares) found in the moving
phase.   Presumably, during transverse lattice motion, these quenched defects
slip, possibly reducing
the transverse critical current below that
expected from an elastic theory which does not allow such slippage.
All these dynamic phases will probably manifest themselves more
clearly in three dimensions.  
In a BiSrCaCuO-type materials, observation
by neutron diffraction\cite{cubitt93} of a field-induced
disordering transition at low temperatures was interpreted in terms of a 
topological phase transition from an ordered line lattice to a
decoupled glass phase\cite{ryu96_2}.  The  glass phase has
quasi-2D topological defects at low $T$, which may have $IV$
similar to those described here.
Thus, by driving the vortices with a high 
$J$ {\em at varying fields}, and carrying out both resistivity and Hall 
measurements, one may be able to probe the moving lattice phase and its
lower critical dimension.

As this work was being completed, we became aware of work by Moon 
{\em et al.} on the similar model\cite{moon96}. 
The authors gratefully acknowledge 
financial support by NSF Grant DMR94-02131 and the
Midwest Superconductivity Consortium through 
DOE Grant DE-FG02-90ER-45427.

\begin{figure}
\caption [first figure] {Simulated $IV$ characteristics for 
$T/T_m = 0.043$, $0.43$, $0.65$, $0.78$, $0.87$, $0.96$, $1.087$. 
Inset: J-T phase diagram constructed from the IV results and analyses of 
$\Psi_6$ (Fig.\ 2). 
The solid line $(J/J_p = 1.2 + 1.2/(1 - T/T_m))$ 
and the broken lines are guides to the eye.
Acronyms defined in the text.}
\label{figone}
\end{figure}

\begin{figure}
\caption [second figure] {Hexatic order parameter $\Psi_6$ (broken line) and 
differential resistance $dV/dI$ for $T/T_m = 0.043.$  
Inset: vortex trajectories at points $b$ and $d$ in the plastic
regime taken over $1000 t_0$. The dots represents those pinned vortices
which remained static over this interval.
Lower panel shows the time averaged g(r)
function for $J/J_p:$ $0.04, 0.4,$ and  $1.98$; they are marked $a$ (PHVG), 
$b$ (PFVL), and $c$ (MHVG).}
\label{figtwo}
\end{figure}

\begin{figure}
\caption [third figure] {Simulated $dV/dI$ for 
$T/T_m = 0.78, 0.87, 0.96, 1.087$ and $1.30$ near melting.
Inset shows $\Psi_6$. 
The crossing of all the
$dV/dI$ curves near $J/J_p \sim 0.12$ coincides with the dip in $\Psi_6$.}
\label{figthree}
\end{figure}

\begin{figure}
\caption [fourth figure] {Transverse voltage induced by a transverse force
$\hat{z}\times J_t \parallel \hat{\bf x}$.  $J_t$ is turned on after 
a steady state is reached under the dominant force 
$\hat{z}\times J_{\ell}\parallel \hat{\bf y}$. Both $T/T_m = 0.043$ (solid) and 
$0.43$ (broken) cases are shown. Inset: snapshot of vortices (points) 
for $\theta = 0, T/T_m = 0.043$.  A pattern of disclinations with coordination 
numbers $z=8$(thick circle), $=7$(circle) and $=5$(square) is also shown.}
\label{figfour}
\end{figure}

\break


\begin{thebibliography}{10}

\bibitem{larkin}
A. I. Larkin, {\it Sov. Phys. JETP} {\bf 31}, 784 (1970);
A. I. Larkin and Y. N. Ovchinnikov, {\it ibid} {\bf 34}, 651 (1972).

\bibitem{statics}
J.-P. Bouchaud {\it et al.},{\em Phys. Rev. B} {\bf 46}, 14686 (1992);
T. Nattermann,{\em Phys. Rev. Lett.} {\bf 64}, 2454 (1990);T. Giamarchi and P.~L. Doussal,
{\em ibid} {\bf 72}, 1530 (1994);T. Giamarchi and P.~L. Doussal, {\em Phys. Rev. B} {\bf 52}, 1242
(1995);E.~M. Chudnovsky,{\em ibid} {\bf 40}, 11355 (1989).

\bibitem{schmid73} A. Schmid and W. Hauger, {\it J. Low Temp. Phys.} {\bf 11},
667 (1973).

\bibitem{jensen88} H. J. Jensen {\it et al.}, {\it Phys. Rev. Lett.} {\bf 60}, 1676 (1988).

\bibitem{shi91} A.-C. Shi and A. J. Berlinsky, {\it Phys. Rev. Lett.} {\bf 67}, 1926 (1991).

\bibitem{jensen96}N. Gr{\o}enbech-Jensen {\it et al.}, {\it Phys. Rev.
Lett.} {\bf 76}, 2985 (1996).

\bibitem{koshelev94} A. E. Koshelev and V. M. Vinokur, {\it Phys. Rev. Lett.}
{\bf 73}, 3580 (1994).

\bibitem{giamarchi96} T. Giamarchi and P. L. Doussal, {\it Phys. Rev. Lett.}
{\bf 76}, 3408 (1996).

\bibitem{cha94} M.-C. Cha and H. A. Fertig, {\it Phys. Rev. B} {\bf 50},
14368 (1994).

\bibitem{falo90} F. Falo {\it et al.}, {\it Phys. Rev. B} {\bf 41},
10983 (1990).

\bibitem{seshadri92}R. Seshadri and R. M. Westervelt, 
{\it Phys. Rev. B} {\bf 46}, 5142 (1992);
(1992); {\bf 46}, 5150 (1992).

\bibitem{fisher85}D. Fisher, {\it Phys. Rev. B} {\bf 31}, 1396 (1985);
A. Middleton, {\it Phys. Rev. Lett.} {\bf 68}, 670 (1992).

\bibitem{balents95} L. Balents and M. P. A. Fisher, {\it Phys. Rev. Lett.} {\bf 75}, 4270 (1995).

\bibitem{shobo95}S. Bhattacharya and M. J. Higgins, {\it Phys. Rev. Lett.}
{\bf 70}, 2617 (1993); S. Bhattacharya and M. J. Higgins, {\it Phys. Rev. B} 
{\bf 49}, 10005 (1994); A. C. Marley {\it et al}, {\it Phys. Rev. Lett.} {\bf 74}, 
3029 (1995).

\bibitem{yaron95} U. Yaron {\it et al}, {\it Nature} {\bf 376}, 753 (1995).

\bibitem{monica96} M. C. Hellerqvist {\it et al.},
{\it Phys. Rev. Lett.} {\bf 76}, 4022 (1996).


\bibitem{ryu96} Seungoh Ryu and D. Stroud, 
\newblock {\em Phys. Rev. B}, in press (1996).

\bibitem{ryu92}
S. Ryu {\it et al},
\newblock {\em Phys. Rev. Lett.} {\bf 68}, 710 (1992).

\bibitem{hexdef}
After Delaunay triangulation, the local hexatic order parameter is 
evaluated from $\Psi_6 = |{ 1\over N} < \sum_{i,j'} {1 \over c_{i}} 
\exp [ i 6 \theta_{ij}] > | $ where $N$ is the total number of pancakes, 
$c_{i}$ the vortex coordination number, and $\theta_{ij}$ is the
bond angle between neighboring vortices relative to a fixed
direction $\hat{x}.$

\bibitem{fisher80} B. I. Halperin and D. R. Nelson, {\it Phys. Rev. Lett.} {\bf 41},
121 (1978); A. P. Young, {\it Phys. Rev. B}, {\bf 19} 1855
(1979); B. A. Huberman and S. Doniach, {\it Phys. Rev. Lett.} {\bf 43}, 950 (1979);D.
S. Fisher, {\it Phys. Rev. B} {\bf 22}, 1190 (1980).

\bibitem{pinnedlattice}
Representative movies showing this behavior can be found at
http://www.physics.ohio-state.edu:80/\~\/ ryu/moge.html.

\bibitem{elsewhere} S. Ryu {\it et al.} (unpublished)

\bibitem{cubitt93} R. Cubitt {\it et al.}, {\it Nature} {\bf 365}, 407 (1993).

\bibitem{ryu96_2}
S. Ryu {\it et al.}, {\em SPIE Proceedings} {\bf
2157}, 12, (1994); Seungoh Ryu, PhD thesis, Stanford
University, (1995); Seungoh Ryu {\it et al.} {\it preprint} (1996).
See also M. J. P. Gingras and D. A. Huse, {\it preprint} (1996).

\bibitem{moon96} K. Moon {\it et al.}, 
{\it preprint} (1996).

\end{thebibliography}
\end{document}